\documentclass[%
    aps,
    prl,
    reprint,
    preprintnumbers,
    superscriptaddress
]{revtex4-2}

\usepackage{amsmath}
\usepackage{amsfonts}
\usepackage{amssymb}
\usepackage{mathrsfs}
\usepackage{mathdots}
\usepackage{mathtools}

\usepackage{xparse}
\usepackage{xstring}
\usepackage{xspace}
\usepackage{xcolor}

\usepackage{array}
\usepackage{enumitem}

\usepackage[colorlinks=true]{hyperref}

\usepackage{tikz}
\usepackage{standalone}

\makeatletter
\DeclareRobustCommand\onedot{\futurelet\@let@token\@onedot}
\newcommand\@onedot{\ifx\@let@token.\else.\null\fi\xspace}
\newcommand{\eg}{\emph{e.g}\onedot} 
\newcommand{\ie}{\emph{i.e}\onedot}

\newcommand{\vs}{\emph{vs}\onedot}
 
\makeatother

\renewcommand{\geq}{\geqslant}

\renewcommand{\Re}{\operatorname{Re}}
\renewcommand{\Im}{\operatorname{Im}}
\AtBeginDocument{\renewcommand{\bar}[1]{\overline{#1}}}
\AtBeginDocument{\renewcommand{\hat}[1]{\widehat{#1}}}
\newcommand{\wave}[1]{\widetilde{#1}}

\newcommand{\nn}{\nonumber}

\newcommand{\pseries}{1+i\, \RR}
\newcommand{\set}[1]{\{ #1 \}}

\newcommand{\zb}{\bar{z}}
\newcommand{\RR}{\mathbb{R}}
\newcommand{\CC}{\mathbb{C}}
\newcommand{\NN}{\mathbb{N}}

\newcommand{\oo}{\infty}
\newcommand{\cA}{\mathcal{A}}

\newcommand{\peq}{\phantom{{}={}}}

\providecommand{\ket}[1]{{| {#1} \rangle}}
\renewcommand{\ket}[1]{{| {#1} \rangle}}

\newcommand{\wb}{\bar{w}}

\newcommand{\transfer}{\mathcal{T}}
\newcommand{\correlator}{\mathcal{F}}
\newcommand{\sllie}{\mathfrak{sl}}
\newcommand{\solie}{\mathfrak{so}}
\newcommand{\measureOfA}{\Omega}

\newcommand{\measureOfC}{\xi}
\newcommand{\cN}{\mathcal{N}}

\newcommand{\shadow}[1]{\wave{#1}}

\newcommand{\bDelta}{\boldsymbol{\Delta}}
\newcommand{\bJ}{\boldsymbol{J}}
\newcommand{\bz}{\boldsymbol{z}}

\newcommand{\id}{\text{\usefont{U}{bbold}{m}{n}1}}
\MakeRobust{\id}

\begin{document}

\title{Celestial Optical Theorem}

\author{Reiko Liu} 
\email{reiko@tsinghua.edu.cn}
\affiliation{Yau Mathematical Sciences Center (YMSC), Tsinghua University, Beijing, 100084, China}

\author{Wen-Jie Ma}
\email{wenjie.ma@simis.cn}
\affiliation{Center for Mathematics and Interdisciplinary Sciences, Fudan University, Shanghai, 200433, China}
\affiliation{Shanghai Institute for Mathematics and Interdisciplinary Sciences (SIMIS), Shanghai, 200433, China}

\begin{abstract}
\noindent
We derive the celestial optical theorem from the $S$-matrix unitarity, which provides nonperturbative bootstrap equations of conformal partial wave (CPW) coefficients. 
This theorem implies that the imaginary parts of CPW coefficients exhibit a positivity property. 
By making certain analyticity assumptions and using the celestial optical theorem, we derive nonperturbative constraints concerning the analytic structure of CPW coefficients. 
We discover that the CPW coefficients of four massless particles must and can only have simple poles located at specific positions. 
The CPW coefficients involving massive particles exhibit double-trace poles, indicating the existence of double-trace operators in nonperturbative CCFT.
It is worth noting that, in contrast to AdS/CFT, the conformal dimensions of double-trace operators do not have anomalous dimensions.
\end{abstract}

\maketitle

\section{Introduction}

Celestial holography connects four-dimensional quantum gravity in asymptotically flat spacetime to a putative two-dimensional celestial conformal field theory (CCFT) \cite{Pasterski:2016qvg,Pasterski:2017kqt,Strominger:2017zoo,Raclariu:2021zjz,Pasterski:2021rjz,Pasterski:2021raf,McLoughlin:2022ljp}.
In this duality, the conformal correlators in the boundary CCFT, known as celestial amplitudes, are obtained by expressing the bulk scattering amplitudes in terms of the conformal basis.

In CFT, conformal correlators can be expanded into conformal partial wave (CPW) coefficients through the CPW expansion. Conversely, CPW coefficients can be extracted from conformal correlators using the inversion formula. The dynamical information, such as operator spectra and three-point coefficients, is directly related to the analytic structure of the CPW coefficients. 
As conformal correlators in CCFT, celestial amplitudes also share this CPW expansion, as shown by perturbative examples in \cite{Lam:2017ofc,Garcia-Sepulveda:2022lga,Atanasov:2021cje,Chang:2021wvv,Fan:2021pbp,Fan:2021isc,Fan:2022kpp,Chang:2022jut,Chang:2023ttm,Fan:2023lky,Himwich:2023njb,Liu:2024lbs}.

\begin{figure}[htbp!]
\centering
\includegraphics[width=0.8\linewidth]{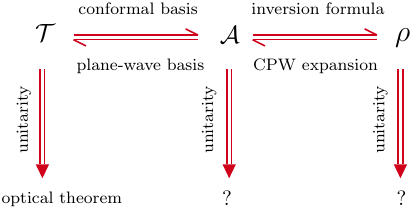}
\caption{%
    The relation between scattering amplitude $\transfer$, celestial amplitude $\cA$ and CPW coefficient $\rho$.
}
\label{fig:main_idea}
\end{figure}

All the three objects - scattering amplitude, celestial amplitude, and CPW coefficient - encode the same physical information. For the scattering amplitude, the (generalized) optical theorem, as a direct consequence of bulk unitarity, serves as the starting point of $S$-matrix bootstrap, see \cite{Eden,Correia:2020xtr,Mizera:2023tfe} and the references therein.
This naturally raises the question (see Figure \ref{fig:main_idea}): 
\begin{quote}
    How does bulk unitarity constrain celestial amplitudes or CPW coefficients in CCFT?
\end{quote}
This question was explored in \cite{Lam:2017ofc,Law:2020xcf,Chang:2021wvv,Chang:2022jut,Iacobacci:2022yjo,Garcia-Sepulveda:2022lga,Ghosh:2022net}.
Particularly, the authors in \cite{Lam:2017ofc} discovered that for a particular tree-level exchange diagram, the optical theorem relates the CPW coefficient to the three-point coefficients.
However, this relation has only been shown at the tree-level, and the nonperturbative implications of bulk unitarity for CCFT remain unknown.

Here, we answer this question by expanding the optical theorem with conformal basis and CPWs. The result, which we call the celestial optical theorem, provides nonperturbative relations between lower- and higher-point CPW coefficients, and can serve as the bootstrap equations of CCFT.

By the celestial optical theorem, we find that for elastic scattering, the imaginary parts of the CPW coefficients with appropriate conformal dimensions are nonnegative. 
Moreover, we derive nonperturbative constraints on the analytic structure of the CPW coefficients. 
Our analysis reveals that for elastic scattering of two massless particles, the CPW coefficients must and can only have simple poles located at specific positions. 
While if involving massive particles, the CPW coefficients contain double-trace poles, suggesting the existence of double-trace operators in CCFT, even in nonperturbative scenarios.

\section{Background}

\textbf{Generalized optical theorem.} 
We consider scattering process of bosonic particles in $4d$ Minkowski spacetime. A generic $n_{K}$-particle state is denoted as $|K\rangle\equiv|\boldsymbol{\alpha}_K,\boldsymbol{p}_K,\boldsymbol{\ell}_K\rangle$, where the bold symbols $\boldsymbol{\alpha}_K$, $\boldsymbol{p}_K$ and $\boldsymbol{\ell}_K$ are the collection of particle species, on-shell momenta and bulk spins of individual particles respectively.
The completeness relation of the Hilbert space is 
\begin{align}
    \label{eq:completeness_relation}
    \id=\sum_{K}\int d\Pi_{K} |K\rangle \langle K|
    \, ,
\end{align}
where $\int d\Pi_{K}$ denotes the on-shell integral together with polarization sums and appropriate permutation factors for identical particles.

The unitarity of the $S$-matrix is equivalent to $iT^{\dagger}-iT=T^{\dagger}T=TT^{\dagger}$ with $S=\id+iT$. From unitarity and completeness \eqref{eq:completeness_relation}, for a scattering process $I\to F$, the optical theorem from $S^{\dagger}S=1$ provides a nonlinear relation of scattering amplitudes:
\begin{align}
    \label{eq:generalized_optical_theorem_QFT}
    i(\transfer_{FI})^{*}
    -
    i\transfer_{IF}
    =
    \sum_{K}\int d\Pi_{K}
    (\transfer_{FK})^{*}
    \transfer_{IK}
    \, ,
\end{align}
where $\transfer_{IF}\equiv\langle F|T|I\rangle$ and similar for others. 
Here and in the following we use $(\cdots)^{*}$ to denote the complex conjugate of $(\cdots)$.

\textbf{Celestial holography.}
In celestial holography, to manifest the conformal symmetry $\sllie(2,\CC)\simeq \solie(3,1)$ on the boundary celestial sphere, the conformal basis 
\begin{align}
    \label{eq:conformal_basis}
    |K\rangle_{\partial}\equiv|\boldsymbol{\alpha}_K,\boldsymbol{\Delta}_K,\boldsymbol{J}_K, \boldsymbol{z}_K\rangle
\end{align}
is introduced for a multi-particle state.
As previously, the bold symbols $\boldsymbol{\Delta}_K$, $\boldsymbol{J}_K$, and $\boldsymbol{z}_K$ denote the collection of conformal dimensions, conformal spins and celestial coordinates respectively.
The $\zb$-dependence is omitted when there is no ambiguity.

The transition matrix from $\ket{K}$ to $\ket{K}_{\partial}$ is called the conformal primary wavefunction, denoted as $\phi$ \cite{Pasterski:2016qvg}. 
Similar to the plane-waves, the conformal primary wavefunctions provide a complete basis to the solutions of the equation of motion. While the plane-waves manifest the translation symmetry, the conformal primary wavefunctions transform covariantly under the conformal group.

As an example, with the parametrization of the null momentum $p=\omega(1+w\wb,w+\wb,-i(w-\wb),1-w\wb)$, the conformal primary wavefunction of a massless scalar is \cite{Pasterski:2016qvg,Pasterski:2017kqt}
\begin{equation}
    \phi_{\Delta}(z,p)
    =
    \frac{1}{4}
    \omega^{\Delta-2}
    \delta^{(2)}(z-w)
    \, .
\end{equation}
For massive and spinning particles, the construction and properties of conformal primary wavefunctions can be found in \eg \cite{Pasterski:2016qvg,Pasterski:2017kqt,Law:2020tsg,Pasterski:2021fjn,Chang:2022seh,Camporesi:1994ga,Costa:2014kfa}.

Given a momentum-space scattering amplitude $\transfer_{IF}$, the celestial amplitude $\mathcal{A}_{IF}$ is defined by expanding it with respect to the conformal basis \cite{Pasterski:2016qvg,Pasterski:2017kqt}. The celestial amplitude $\cA_{IF}$ transforms as a boundary conformal correlator with coordinates $\bz_{I}$, $\bz_F$ and conformal weights $(\bDelta_I,\bJ_I)$, $(\bDelta_F,\bJ_F)$, \textit{i.e.},
\begin{align}
    \label{eq:abbreviation_A}
    \cA_{IF}\equiv
    \cA^{\boldsymbol{\Delta}_I,\boldsymbol{J}_I|\boldsymbol{\Delta}_F,\boldsymbol{J}_F}_{\boldsymbol{\alpha}_I| \boldsymbol{\alpha}_F}(\boldsymbol{z}_{I}|\boldsymbol{z}_F)
    \, .
\end{align}
For example, the celestial amplitude of scalars takes the form
\begin{align}
    \label{eq:CA}
    &\mathcal{A}_{IF}
    =
    \bigg(\prod_{a=1}^{n_I+n_F} \int\frac{d^3p_a}{p_a^0}\phi_{\Delta_a}(z_a,p_a)\bigg)
    \transfer_{IF}
    \, .
\end{align}
Since the transformation is invertible, the celestial amplitude $\cA_{IF}$ captures the same physical information as the scattering amplitude $\transfer_{IF}$.

\begin{figure}[htbp!]
\centering
\includegraphics[width=\linewidth]{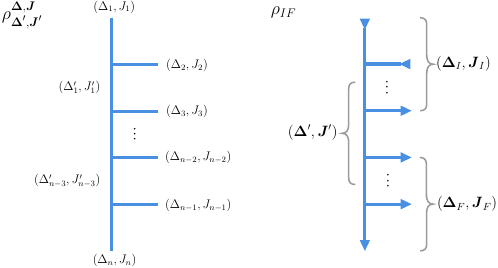}
\caption{%
    Comb-channel CPW coefficients.
    The left is the CPW coefficient $\rho^{\bDelta,\bJ}_{\bDelta',\bJ'}$ in \eqref{eq:conformal_partial_wave_expansion}.
    The right is the CPW coefficient $\rho_{IK}$ associated to $\cA_{IF}$, and the arrows denote the direction of incoming/outgoing.
}
\label{fig:comb-channel}
\end{figure}

\textbf{Conformal partial wave expansion.}
As mentioned before, the CPWs $\Psi$ provide a complete basis for expanding conformal correlators \cite{Ferrara:1972uq,Dobrev:1976vr,Dobrev:1977qv,SimmonsDuffin:2012uy,Simmons-Duffin:2017nub,Karateev:2018oml,Liu:2018jhs,Kravchuk:2018htv,Chen:2020vvn,Chen:2022cpx,Chen:2022jhx}. They transform covariantly under the conformal group and are single-valued solutions of conformal Casimir equations. We leave the useful properties of CPWs in the supplementary material.
With the CPWs, the four-point conformal correlator $\correlator^{\bDelta,\bJ}(\bz)$ can be decomposed as
\begin{equation}
    \label{eq:CPW_expansion}
    \correlator^{\bDelta,\bJ}(\bz)
    =
    \sum_{J'=-\oo}^{+\oo}
    \int_{1}^{1+i\infty}\!\!\frac{d\Delta'}{\mu(\Delta',J')}
    \rho^{\bDelta,\bJ}_{\Delta',J'} \Psi^{\bDelta,\bJ}_{\Delta',J'}(\bz)
    \, ,
\end{equation}
where $\rho$ is called the CPW coefficient and  
\begin{equation}
    \mu(\Delta,J)=\frac{4\pi^{4}i}{J^2-(\Delta-1)^2}
\end{equation}
is the Plancherel measure. 
Generalizing to $n$-point, $\correlator$ can be decomposed into the CPW coefficient $\rho$ in the comb-channel \cite{Alkalaev:2015fbw,Rosenhaus:2018zqn,Parikh:2019ygo,Goncalves:2019znr,Jepsen:2019svc,Parikh:2019dvm,Fortin:2019zkm,Fortin:2020yjz,Anous:2020vtw,Haehl:2021tft,Fortin:2020bfq,Hoback:2020pgj,fortin2020all,Buric:2020dyz,Hoback:2020syd,Poland:2021xjs,Buric:2021ywo,Buric:2021ttm,Buric:2021kgy,fortin2022feynman,Fortin:2023xqq} (see the left panel of Figure \ref{fig:comb-channel}):
\begin{align}
    \label{eq:conformal_partial_wave_expansion}
    &\correlator^{\bDelta,\bJ}(\bz)
    =
    \sum_{\bJ'}\int d\bDelta'\,
    \rho^{\bDelta,\bJ}_{\bDelta',\bJ'}
    \Psi^{\bDelta,\bJ}_{\bDelta',\bJ'}(\boldsymbol{z})
    \, .
\end{align}
Here the exchange conformal weights are denoted by primed symbols, \eg $\bDelta'\equiv \set{\Delta'_{1},\dots,\Delta'_{n-3}}$. The vectorial sum and integral are the $(n-3)$-fold version of the ones in \eqref{eq:CPW_expansion}.

In particular, as a boundary conformal correlator, 
we denote the CPW coefficient associated to the celestial amplitude $\cA_{IF}$ as $\rho_{IK}$. Explicitly, $\rho_{IK}$ depends on the $(n_{I}+n_{F})$ external conformal weights $(\bDelta_I,\bJ_I)$, $(\bDelta_F,\bJ_F)$ and $(n_{I}+n_{F}-3)$ exchange ones $(\bDelta',\bJ')$, see the right panel of Figure \ref{fig:comb-channel}.

We emphasize that due to the unitarity condition of the Euclidean conformal group, the conformal dimensions in conformal primary wavefunctions, celestial amplitudes and CPW coefficients are initially restricted on the principal series $\pseries$. In practice they can be analytically continued to a larger domain of holomorphy, and we will adopt this prescription by default.

\section{Celestial Optical Theorem}

In this section, we present the celestial optical theorem, which is a consequence of the bulk unitarity and the completeness relation of the Hilbert space \eqref{eq:completeness_relation}.
In the rest of this letter, we will focus on $2$-to-$2$ scattering $I\to F$, \ie $n_I=n_F=2$ and
\begin{align}
\begin{split}
    &|I\rangle_{\partial}= |\alpha_1,\alpha_2,\Delta_1,\Delta_2,J_1,J_2,z_{1},z_{2}\rangle
    \, ,
    \\
    &|F\rangle_{\partial}= |\alpha_3,\alpha_4,\Delta_3,\Delta_4,J_3,J_4,z_{3},z_{4}\rangle
    \, .
\end{split}
\end{align}
Moreover, for a boundary state $|K\rangle_{\partial}$ in \eqref{eq:conformal_basis}, we define its hatted conjugation $|\hat{K}\rangle_{\partial}$ as
\begin{align}
    \label{eq:hatted_conjugate}
   \hat{}\;:|K\rangle_{\partial}\mapsto |\hat{K}\rangle_{\partial}\equiv|\boldsymbol{\alpha}_K,\boldsymbol{\Delta}^*_K,-\boldsymbol{J}_K,\boldsymbol{z}_K\rangle
   \, .
\end{align}
Here $\boldsymbol{\Delta}^{*}_K$ is a collection of conformal dimensions whose elements are complex conjugate of those in $\boldsymbol{\Delta}_K$. 
As the derivation is standard but tedious, we leave it to the supplementary material. The main idea contains the following two steps. 

\textbf{Step I.}
Similar to the derivation of the optical theorem, we use the completeness relation of the conformal basis to expand the optical theorem \eqref{eq:generalized_optical_theorem_QFT} into celestial amplitudes:
\begin{align}
    \label{eq:optical_A}
    &
    \!\!
    i(\cA_{\hat{F}\hat{I}})^{*}
    -
    i\cA_{IF}
    =
    \sum_{K}
    \int\! d\measureOfA_{K}
    \!
    \int\! d^2\boldsymbol{z}_{K} 
    (\cA_{\hat{F}K})^{*} \cA_{IK}
    \, .
\end{align}
Here $\cA_{\hat{F}\hat{I}}$, depending on the hatted boundary states $|\hat{I}\rangle_{\partial}$ and $|\hat{F}\rangle_{\partial}$, is an abbreviation as in \eqref{eq:abbreviation_A}. The integral over $\measureOfA_{K}$ is
\begin{align}
    \label{eq:int_OmegaK}
    \int d\measureOfA_{K}
    =
    \frac{1}{S_{K}}
    \prod_{a=1}^{n_K}
    \sum_{J_{a}\in L(\ell_{a})}
    \int_{1-i\infty}^{1+i\infty}\frac{d\Delta_{a}}{\cN(\Delta_{a},J_{a},\ell_{a})}
    \, ,
\end{align}
where the symmetry factor $S_{K}$ takes into account identical intermediate particles.
If the $a$-th particle in $K$ is massive with mass $m$ and bulk spin $\ell$, then
the set $L(\ell)$ and the factor $\cN(\Delta,J,\ell)$ are
\begin{align}
\begin{split}
    &L(\ell)=\set{-\ell,-\ell+1,\cdots,\ell-1,\ell}
    \, ,
    \\
    &\cN(\Delta,J,\ell)=
    \frac{
        (-1)^{J+1 } 2^{6+\ell} \pi^{6} i\,    (\ell-J)! (\ell+J)!
    }{ 
       m^{2}  (\Delta+J-1) (\Delta-J-1) (2 \ell)!
    }
    \, .
\end{split}
\end{align}
If the $a$-th particle in $K$ is massless with bulk spin $\ell$, then $L(\ell)$ and $\cN(\Delta,J,\ell)$ are
\begin{align}
\begin{split}
    &L(\ell)=\set{-\ell,\ell}
    \, ,
    \\
    &\cN(\Delta,J,\ell)=
    2^{\ell+3} \pi^4 i
    \, .
\end{split}
\end{align}

\begin{figure}[htbp!]
\centering
\includegraphics[width=\linewidth]{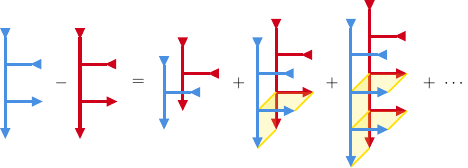}
\caption{%
    Illustration of celestial optical theorem \eqref{eq:celestial_optical:S+S=1}. We use the red color denoting complex conjugation.
    The pairs of red and blue arrows connected by yellow plaquettes represent the variables that should be integrated out.
}
\label{fig:optical_rho}
\end{figure}

\textbf{Step II.}
The relation \eqref{eq:optical_A} serves as a nonperturbative bootstrap equation of celestial amplitudes.
We further expand it into CPW coefficients and reach the following celestial optical theorem:
\begin{align}
    \nn
    &
    i(\rho_{\hat{F}\hat{I}})^{*}
    -
    i\rho_{IF}
    =
    \sideset{}{'}\sum_{K}\measureOfC_{K} 
    \Big(
        (C_{\hat{F}K})^{*}
        C_{IK}
        +
        (K\leftrightarrow \shadow{K})
    \Big)
    \\
    &\peq
    +
    \sideset{}{''}\sum_{K}
    \sum_{\bJ'_{K}}\int d\bDelta'_{K}
    \int d\measureOfA_{K}
    \, 
    (\rho_{\hat{F}K})^{*}
    \rho_{IK}
    \, .
    \label{eq:celestial_optical:S+S=1}
\end{align}
This equation can be illustrated by Figure \ref{fig:optical_rho}.

\begin{figure}[htbp!]
\centering
\includegraphics[width=\linewidth]{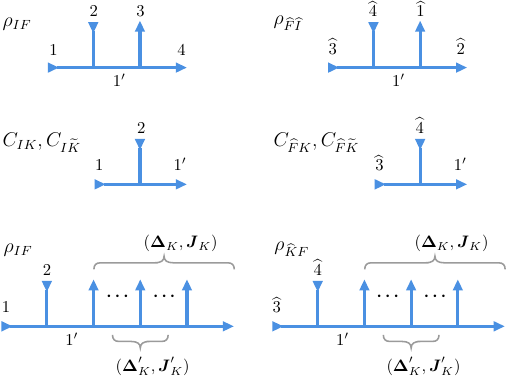}
\caption{%
    Dependence of each term in \eqref{eq:celestial_optical:S+S=1} on external and exchange conformal dimensions and spins. 
    Here $1 $ denotes $(\Delta_{1},J_{1})$, $\hat{3}$ denotes $(\Delta^{*}_{3},-J_{3})$, $1'$ denotes $(\Delta'_{1},J'_{1})$, and similar for others.
    Notice that $(\bDelta'_K, \bJ'_K)$ denotes the collection of exchange conformal weights without the first one $(\Delta'_{1},J'_{1})$.
}
\label{fig:rho}
\end{figure}

Now we explain this equation in detail. 
For each term in \eqref{eq:celestial_optical:S+S=1}, the dependence on conformal weights is summarized in Figure \ref{fig:rho}.
The free variables are the external conformal weights $(\Delta_{i},J_{i})$ for $i=1,2,3,4$ and the exchange one $(\Delta'_1, J'_1)$.

On the right side of \eqref{eq:celestial_optical:S+S=1}, the primed sum accounts for contributions from intermediate single-particle states, while the double-primed sum includes contributions from multi-particle states.
$C_{IK}$ is the three-point coefficient of the $2$-to-$1$ scattering $I\to K$, and $C_{I\shadow{K}}$ is the one with the outgoing particle $K$ expanded by the shadow conformal basis \cite{Pasterski:2017kqt,Chang:2022jut,Chang:2022seh,Furugori:2023hgv,Liu:2024lbs}.
The factor $\xi_{K}(\Delta'_{1},J'_{1})$ is 
\begin{equation}
    \xi_{K}(\Delta,J)=
    \begin{cases}
        \frac{\mu(\Delta,J)}{
        \cN(\Delta,J,\ell_{K})
        }, &\text{if }\, J\in L(\ell_{K})
        \, ,
        \\
        0, & \text{if }\, J\notin L(\ell_{K})
        \, .
    \end{cases}
\end{equation}
The integral $\int d\measureOfA_{K}$ is defined in \eqref{eq:int_OmegaK} and $\sum_{\bJ'_{K}}\int d\bDelta'_{K}$ is the $(n_K-2)$-fold version of the ones in \eqref{eq:CPW_expansion}.
Here, $(\bDelta'_{K},\bJ'_{K})$ is the collection of exchange conformal weights excluding the first one $(\Delta'_1, J'_1)$, see Figure \ref{fig:rho}.

The relation \eqref{eq:celestial_optical:S+S=1} serves as a bootstrap equation in the boundary CCFT imposed by the bulk unitarity $S^{\dagger}S=1$, and we expect numerous physical information can be extracted from it.
Additionally, another celestial optical theorem can be similarly derived from $SS^{\dagger}=1$, which we have included in the supplementary material.

\textbf{Celestial optical theorem \vs optical theorem.} 
The optical theorem \eqref{eq:generalized_optical_theorem_QFT} and the celestial optical theorem \eqref{eq:celestial_optical:S+S=1} are equivalent up to the change of scattering basis, and both of them serve as bootstrap equations. 
In the conventional $S$-matrix program, to utilize the optical theorem, the $2$-to-$2$ scattering amplitudes are further expanded by Legendre/Jacobi polynomials for scalar/spinning particles.
However, for scattering amplitudes with more than four particles there is currently no such polynomial basis. This lack of knowledge makes such amplitudes challenging to handle.
Moreover, the optical theorem involves on-shell vectorial integrals which are difficult to manipulate. 
In contrast, there exists a natural basis in CCFT - the conformal partial waves - to expand higher-point celestial amplitudes. 
This enables us to obtain scalar equations of CPW coefficients containing only Mellin-Barnes integrals. 

\section{Applications}

We assume that the bulk $S$-matrix is unitary so that the celestial optical theorem \eqref{eq:celestial_optical:S+S=1} holds, and will study properties of the CPW coefficient.
We focus on the elastic scattering process, \ie the 3-rd (4-th) outgoing particle has the same species as the 1-st (2-nd) incoming particle. In this case, for fixed external conformal spins $J_{i}$ with $i=1,2,3,4$ and exchange one $J'_{1}$, the CPW coefficient $\rho_{IF}$ is a function of external conformal dimensions $\Delta_{i}$ and exchange one $\Delta'_{1}$. 
For clarity we relabel the exchange conformal weights as $(\Delta'_{1},J'_{1})\equiv(\Delta,J)$ and the CPW coefficient as $\rho_{IF}\equiv \rho_{J}^{J_{i}}(\Delta;\Delta_{i})$.

\begin{figure}[htbp!]
\centering
\includegraphics[width=0.8\linewidth]{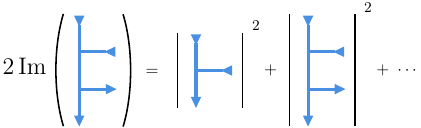}
\caption{Illustration of positivity.}
\label{fig:positivity}
\end{figure}

\textbf{Positivity.} 
Setting $(\alpha_1,\Delta_1,J_1)=(\alpha_3,\Delta_{3}^{*},-J_3)$ and $(\alpha_2, \Delta_{2},J_2)=(\alpha_4,\Delta_{4}^{*},J_4)$ in \eqref{eq:celestial_optical:S+S=1}, each term on the right side is manifestly nonnegative (see also Figure \ref{fig:positivity}), hence we obtain the following positivity property:
%
\begin{align}
    \label{eq:Imrho>0}
    \Im\rho^{J_1,J_2,-J_1,-J_2}_{J}(\Delta;\Delta_1,\Delta_2,\Delta_1^{*},\Delta_2^{*})\geq 0
\end{align}
for $\Delta\in 1+i\, \RR$.

If the putative CCFT corresponds to a unitary bulk theory, the CPW coefficients must satisfy this positivity condition \eqref{eq:Imrho>0}. We mention that similar positivity conditions have also been established in \cite{Chang:2021wvv}. There, the authors found that if the external conformal dimensions are real, the imaginary part of the four-point celestial amplitude can be positively expanded by the Poincare partial waves.

\textbf{Analyticity.} 
As mentioned previously, $\rho_{J}^{J_{i}}(\Delta;\Delta_{i})$ is initially defined for the conformal dimensions $\Delta$ and $\Delta_{i}$ on the principal series $\pseries$. In practice it can be extended to a meromorphic function on a larger domain. Particularly, when the corresponding conformal correlator admits a convergent conformal block expansion, we may naturally assume that $\rho$ is meromorphic for $\Delta\in \CC$. Besides, perturbative examples suggest that $\rho$ is also meromorphic for $\Delta_{i}$.
We make the above more rigorous into the following assumptions:
\begin{enumerate}
    \item \label{assumption:CFT}
    $\rho_{J}^{J_{i}}(\Delta;\Delta_{i})$ is meromorphic with respect to $\Delta\in \CC$ and decays sufficiently fast to zero as $\Delta\to \oo$;
    \item \label{assumption:meromorphicity} 
    $\rho_{J}^{J_{i}}(\Delta;\Delta_{i})$ is meromorphic with respect to $\Delta_{i}\in \CC^{4}$.
\end{enumerate}

Under the above assumptions and by the technique of complex analysis, we can prove the following properties of $\rho_{J}^{J_{i}}(\Delta;\Delta_{i})$ for generic $\Delta_{i}$ 
\footnote{%
    A property holds for generic $\Delta_{i}\in \CC^{4}$ means that it holds in a dense subset in $\CC^{4}$, but it may not hold on some lower-dimensional analytic subvariety in $\CC^{4}$ by fining tuning of $\Delta_{i}$.
}. The proofs are left in the supplementary material.

\textbf{Property 1.} Exactly one of the following statements is true:
    \begin{enumerate}
        \item for any $J$ and $J_{i}$, $\rho_{J}^{J_{i}}(\Delta;\Delta_{i})=0$;
        \item for $J_{1}=-J_{3}$ and $J_{2}=-J_{4}$, $\rho_{J}^{J_{i}}(\Delta;\Delta_{i})$ contains at least one $\Delta$-pole. 
    \end{enumerate}

\textbf{Property 2.} Given any three-point coefficient 
$C_{IK}$ illustrated in Figure \ref{fig:rho} with $(\Delta'_1,J'_1)\equiv(\Delta,J)$, a pole $\Delta=f(\Delta_{1},\Delta_{2})$ of $C_{IK}$ is also a pole of $\rho_{J}^{J_{i}}(\Delta;\Delta_{i})$ for $J_{1}=-J_{3}$ and $J_{2}=-J_{4}$, if the following conditions hold:
\begin{enumerate}
\item $f$ is meromorphic with respect to $(\Delta_{1},\Delta_{2})\in\CC^{2}$;
\item the equation $\Re f(\Delta_{1},\Delta_{2})=1$ has solutions for $(\Delta_{1},\Delta_{2})\in\CC^{2}$.
\end{enumerate}

\textbf{Property 3.} If the two incoming particles are both massless, exactly one of the statements is true:
    \begin{enumerate}
        \item for any $J$ and $J_{i}$, $\rho_{J}^{J_{i}}(\Delta;\Delta_{i})=0$;
        \item for $J_{1}=-J_{3}$ and $J_{2}=-J_{4}$, $\rho_{J}^{J_{i}}(\Delta;\Delta_{i})$ must and can only contain simple $\Delta$-poles located at 
        \begin{equation}
        \begin{split}
            &
            \Delta=-\Delta_{34}-J_{34}-J+2(n+1)
            \, ,
            \\
            &
            \Delta=\Delta_{34}-J_{34}+J+2(n+1)
            \, ,
            \\
            &
            \Delta=\Delta_{12}+J_{12}-J-2n
            \, ,
            \\
            &
            \Delta=-\Delta_{12}+J_{12}+J-2n
            \, ,
        \end{split}
        \end{equation}
        for some $n\in \NN$. 
        Here $\Delta_{ij}\equiv\Delta_{i}-\Delta_{j}$ and similar for $J_{ij}$.
    \end{enumerate}

\textbf{Comparison to known examples.} The validity of our properties can be verified by examining the known examples of CPW coefficients \cite{Lam:2017ofc,Nandan:2019jas,Chang:2022jut,Chang:2023ttm,Liu:2024lbs}. Particularly, using the results from \cite{Pasterski:2017ylz,Nandan:2019jas}, we find the $\Delta$-poles of CPW coefficients for the MHV amplitudes fall into the set described in the Property 3.

\textbf{Existence of double-trace operators.}
The Property 2 implies that the analytic structure of the four-point CPW coefficient $\rho_{IF}$ is closely related to that of the three-point coefficient $C_{IK}$. 
Here, we provide a concrete example of scalar particles.

We first show the $\Delta$-poles in the scalar three-point coefficient $C_{IK}$ will not be corrected by loop diagrams.
Using \eqref{eq:CA}, any scalar three-point celestial amplitude takes the form as
\begin{align}\label{eq:cA3pt}
    \cA_{IK}=
    \left(\prod_{a=1}^3\!\int\!\frac{d^3p_a}{p^0_a}\phi_{\Delta_a}\!\right)\delta^{(4)}(p_1+p_2-p_3)\mathcal{M}_{IK}
    \, ,
\end{align}
where $\Delta_3\equiv\Delta$ and $\mathcal{M}_{IK}$ is the scattering amplitude $\transfer_{IF}$ with the momentum conservation $\delta$-function dropped off. 
Using Lorentz symmetry and momentum conservation, $\mathcal{M}_{IK}$ only depends on the mass squares and thus can be taken out from the integral.
The remaining integral in \eqref{eq:cA3pt} is exactly the tree-level three-point celestial amplitude without coupling constant, and the $\Delta$-poles can only come from this integral.

Together with Property 2, we conclude that the $\Delta$-poles in the full  $\rho_{IF}$ is directly related to the ones in the tree-level $C_{IK}$. 

As an example, we consider the case that at least one of the two incoming particles in $I$ is massive and there exists a nonvanishing three-point coefficient $C_{IK}$ with $\ell_{K}=0$ \footnote{If the two incoming particles are all massless scalars, the celestial amplitudes do not have proper conformal block expansions \cite{Chang:2022jut,Liu:2024lbs}, and the shadow conformal basis was introduced to rescue this problem \cite{Chang:2022seh}. Then by the Property 3, it can be shown that the CPW coefficients with shadow basis involved only contain double-trace poles.}.
As shown in \cite{Liu:2024lbs}, the tree-level scalar three-point coefficients $C_{IK}$ involving more than two massive scalars have poles at $\Delta=\Delta_1+\Delta_2+2n$ for $n\in \NN$, which implies that the full $C_{IK}$ also have the same poles as we discussed above. 
Then since $\Delta=\Delta_1+\Delta_2+2n$ is meromorphic and $\Re(\Delta_1+\Delta_2+2n)=1$ has solutions for $(\Delta_1,\Delta_2)\in \CC^{2}$, by the Property 2, $\rho_{IF}$ must contain double-trace poles at $\Delta=\Delta_1+\Delta_2+2n$ for $n\in \NN$ nonperturbatively \footnote{%
    In AdS/CFT, the double-trace operators $\mathcal{O}_1(\partial^{2})^{n}\mathcal{O}_2$ appear in the $\mathcal{O}_{1}\times \mathcal{O}_{2}$ OPE.
    Their conformal dimensions are $(\Delta_1+\Delta_2+2n)$ at the tree-level, and will be corrected by loop diagrams.
    The $\Delta$-poles in CPW coefficients with this form are called double-trace poles, and here we adopt this terminology from AdS/CFT.
}.

\section{Discussion}

The celestial optical theorem \eqref{eq:celestial_optical:S+S=1} can serve as a nonperturbative bootstrap equation in celestial holography. 
While we have focused on the $2$-to-$2$ scattering, this equation can be trivially generalized to arbitrary scatterings, resulting in a complete set of bootstrap equations of CPW coefficients.
We have derived several analyticity properties of CPW coefficients, and it would be intriguing to explore numerical methods for solving these equations.

There are two other ingredients in celestial holography - the conformally soft theorem \cite{He:2014laa,Kapec:2016jld} and the $w_{1+\infty}$ symmetry \cite{Strominger:2021mtt}.
We expect that our equations, together with the constraints from the conformally soft theorem and the $w_{1+\infty}$ symmetry, would provide a more precise prescription for celestial bootstrap.

The positivity \eqref{eq:Imrho>0} provides a criterion to discriminate which CFT can be celestial, \ie, corresponding to a unitary bulk theory. Moreover, it also reminds us the cosmological bootstrap \cite{Hogervorst:2021uvp,DiPietro:2021sjt}, where the dS unitarity implies that the CPW coefficient itself is nonnegative, instead of the imaginary part. Recent attempts trying to relate celestial amplitudes with (E)AdS/dS correlators can be found in \cite{Iacobacci:2022yjo,Casali:2022fro,Melton:2023bjw,Melton:2024gyu}.

Another avenue is to explore the implication of the $S$-matrix crossing symmetry on the CPW coefficients. 
The two central ingredients in the modern $S$-matrix bootstrap are unitarity and crossing symmetry, see \eg \cite{Eden,Paulos:2016but,Paulos:2016fap,Paulos:2017fhb,Homrich:2019cbt,Correia:2020xtr,Mizera:2023tfe}, and the former have been encoded in the celestial optical theorem. It would be of great interest to study the celestial counterpart of the bulk crossing symmetry. We expect that the crossing symmetry of $S$-matrix would provide another set of equations on the CPW coefficients. 

\begin{acknowledgments}
    The authors would like to thank Chi-Ming Chang, Ellis Ye Yuan and Tian-Qing Zhu for useful discussions. WJM is supported by the National Natural Science Foundation of China No. 12405082.
\end{acknowledgments}

\bibliography{refs}

\end{document}